\documentclass[aps,preprint,preprintnumbers,nofootinbib,showpacs]{revtex4}
\usepackage{amsmath}
\usepackage{color}

\def\bea{\begin{eqnarray}}
\def\eea{\end{eqnarray}}
\def\lla{\left\langle}
\def\rra{\right\rangle}

\def\ssc{\scriptscriptstyle}
\def\lsim{\mathrel{\raise.3ex\hbox{$<$\kern-.75em\lower1ex\hbox{$\sim$}}} }
\def\gsim{\mathrel{\raise.3ex\hbox{$>$\kern-.75em\lower1ex\hbox{$\sim$}}} }

\begin{document}


\vspace*{0.5in}
\title{A Quantum Space Behind Simple Quantum Mechanics}

\author{\bf Chuan Sheng Chew, Otto C. W. Kong, and Jason Payne
\vspace*{.2in}
}
\email{otto@phy.ncu.edu.tw}

\affiliation { Department of Physics and Center for Mathematics and
Theoretical Physics, National Central University, Chung-li, Taiwan 32054  \\
}

\begin{abstract}
\vspace*{.2in}
In physics, experiments ultimately inform us as to what constitutes 
a good theoretical model of any physical concept: physical space should be no exception. The best picture of
physical space in Newtonian physics is given by the configuration space
of a free particle (or the center of mass of a closed system of
particles). This configuration space (as well as phase space), can be constructed
as a representation space for the relativity symmetry. From the
corresponding quantum symmetry, we illustrate the construction of a
quantum configuration space, similar to that of quantum phase space, and 
recover the classical picture as an approximation 
through a contraction of the (relativity) symmetry and its representations. The quantum Hilbert space reduces into a sum of 
one-dimensional representations for the observable algebra, with the 
only admissible states given by coherent states and position eigenstates 
for the phase and configuration space pictures, respectively.
This analysis, founded firmly on known physics, provides a quantum picture of physical space beyond that of a 
finite-dimensional manifold, 
and provides a crucial first link for any theoretical
model of quantum spacetime at levels beyond simple quantum mechanics. It also suggests looking
at quantum physics from a different perspective.
\end{abstract}
\maketitle

\noindent
{\em Preface for  the Special Issue: ``Planck-Scale Deformations of Relativistic Symmetries''}\\
\indent Our group has been working on a relativity deformation scheme
within the Lie group/algebra framework. This setting has the contraction
process as the reverse of the deformation procedure, and as such it can be
applied to the full physical picture through tracing the contraction
of the relevant representation(s). Both the `quantum Galilean' and 
the classical Galilean symmetries arise within the contraction
limits of the full quantum relativity symmetry. This article focuses
on the simple quantum to classical contraction and discusses a quantum
model of the physical space from this perspective.

\section{Introduction}
Quantum mechanics came into physics after a few hundred years of 
Newtonian mechanics, as the latter failed to describe physics at the 
atomic scale and beyond. In our opinion, however, that quantum 
revolution has not been completed. It was easy to accept the 
mathematical formulation of the theory, but a lot more difficult to 
adopt a fundamental change in our basic perspective. It is not a 
surprise, then, that even the great physicists who created the theory 
kept trying to think and talk about it in terms of Newtonian concepts, 
many of which are really not compatible with quantum mechanics. 
The famous Bohr-Einstein debate, in a way, has never ended; such has 
been the pursuit of a `classical' theory behind quantum mechanics. 
Statements about quantum physics being counter-intuitive, for 
example, are commonly seen and believed by many. We tell our 
students that the quantum world is impossible to make sense of; that 
quantum mechanics gives only probabilistic predictions. The thesis 
presented here is that some, if not all, of those beliefs may simply be 
the result of our reluctance to take the necessary quantum jump in our
fundamental perspective, as well as our indulgence in Newtonian concepts. 
The latter is not really any more intuitive than the modified versions 
suggested by quantum mechanics, only more familiar. A key concept,
and the main focus here, is that of space or position. The perspective
here is that quantum mechanics should be looked at as a dynamical 
theory for physical entities in a space that is really quantum instead 
of classical. Position, as a dynamical variable, is not real-valued 
because a quantum space cannot be modeled on a continuum of 
points as it can be in classical commutative geometry, at least not 
a finite-dimensional one \cite{065}.

The idea of a quantum geometry is certainly not new; however,
here we are talking about a picture of that quantum space completely 
at the level of simple, textbook, so-called non-relativistic quantum 
mechanics. Moreover, we will justify it and illustrate explicitly how 
the classical Newtonian picture is retrieved in the classical 
approximation. The formulation presented here is based on 
relativity symmetries and symmetry contractions.

As said above, and as can hardly be emphasized enough, every precise 
formulation of any physical concept is really only a model - or part of a 
model - of nature. Hence, all such concepts need to have their 
mathematical and physical content re-evaluated as theories develop.
Quantum mechanics as it is to date inherits, with little critical revision,
many Newtonian conceptual notions, while we see that perhaps a lot 
more fundamental changes are called for, even down to the most basic 
one: that of physical space and position within it. The key question then
is how we are going to look at the latter as a feature of the model
instead of just as a background assumption. Instead of thinking about
a theory of mechanics as to be constructed on a model of physical
space, we need to see how the mechanical theory informs us as to
what space is. Only then we can analyze what quantum mechanics says 
about physical space and how that is related to the more familiar
Newtonian picture, which one must be able to retrieve as a limit or
an approximation. Here, relativity symmetry - the Galilean symmetry for 
the case of Newtonian mechanics - is the crucial link. It is as fundamental 
as the assumption of the structure of the physical space itself. It is the set 
of admissible reference frame transformations, hence the symmetry of 
space itself. In fact, both physical space taken as the configuration space 
and as the phase space, at least for the most basic physical system of a 
free particle, should be seen as representations of this symmetry. Recall 
that within the Newtonian  theory the center of mass for any closed system (of particles) 
behaves exactly as a free particle, which illustrates the unbiased 
structure of physical space. The relativity symmetry is therefore central 
to a theory of mechanics. Another good illustration of this point is 
provided by the Poincar\'e symmetry for Einsteinian special relativistic 
physics. The problem, though, is that quantum mechanics has not been 
exactly described as having its own relativity symmetry. We suggest it 
does, as illustrated below.

\section{Quantum Kinematics from a Relativity Symmetry.}
Let us look at the mathematical formulation first, and leave issues with 
the conceptual perspective to be discussed below. With justification for 
the terminology being quite self-evident as the formulation develops, 
we consider a (partial) relativity symmetry for simple quantum mechanics 
as being given by the Lie algebra with the following nonzero commutators
\bea &&
[J_{ij}, J_{hk}] =  i(\delta_{jk} J_{ih} - \delta_{jh} J_{ik} + \delta_{ih} J_{jk} 
- \delta_{ik} J_{jh} ) \;,
\qquad\quad
[X_i,P_j]=i \delta_{ij} I \;,
\nonumber \\ &&
[J_{ij}, P_k] = i(\delta_{jk} P_i - \delta_{ik} P_j) \;,
\qquad\quad
[J_{ij}, X_k] = i(\delta_{jk} X_i - \delta_{ik} X_j) \;,
\eea
with indices going from 1 to 3. We could have included the missing 
generator $H$ with only one nonzero commutator:  $[X_i, H] = -iP_i$.  
The full algebra would then just be the nontrivial $U(1)$ central 
extension of the algebra for the Galilean group, for which the $X_i$ 
are usually denoted by $K_i$ and interpreted as generators for the 
Galilean boosts. In fact, that symmetry has been used as the starting 
point for the quantization of Newtonian particle physics \cite{gq}. The 
$K_i$, as observables, indeed give the (mass times) position, while the 
central extension is what allows for the Heisenberg commutation 
relation. The Hamiltonian $H$ has no role to play in the kinematical
descriptions here, nor is including it much of a problem.  
Note that without $H$ we do have a closed 
subalgebra. We denote by $H_{\!\ssc R}(3)$ the symmetry generated 
by this subalgebra, a three dimensional Heisenberg(-Weyl) symmetry 
with rotations included. As we will illustrate below, representations 
of this symmetry describe quantum space, i.e. the quantum configuration 
space, as well as the phase space, for a quantum `particle' with no spin. 

We start with the coset space representation obtained by
factoring out the $SO(3)$ subgroup. The explicit form of a 
generic infinitesimal transformation is given by
\bea \label{qpsc}
\left(\begin{array}{c}
dp^i  \\ dx^i\\ d\theta \\    0
\end{array}\right) =
\left(\begin{array}{cccc}
 \omega^i_j &  0& 0  &  \bar{p}^i\\
0  & \omega^i_j &  0 &  \bar{x}^i \\
-\frac{1}{2}\bar{x}_j  & \frac{1}{2}\bar{p}_j  &  0 & \bar{\theta} \\
 {0}   & {0} &  0 & 0
\end{array}\right)
\left(\begin{array}{c}
p^j \\ x^j  \\ \theta \\    1
\end{array}\right)
=\left(\begin{array}{c}
\omega^i_j  p^j + \bar{p}^i\\
\omega^i_j  x^j + \bar{x}^i \\ 
  \frac{1}{2}( \bar{p}_j   x^j -\bar{x}_j p^j)  + \bar{\theta} \\   0
\end{array}\right) \;,
\eea
where the real parameters $\omega^i_j$, $\bar{p}^i$, $\bar{x}^i$, 
and $\bar{\theta}$ describe the algebra element 
$-i\left(\frac{1}{2}\omega^{ij} J_{ij}+\bar{p}^i X_i -\bar{x}^i P_i+\bar{\theta}I\right)$.
We will see that the coset space with coordinates $(p^i,x^i,\theta)$ is, in 
a way, the counterpart of the phase space for Newtonian mechanics, 
written as a coset space. The fact that the representation is not unitary, 
however, is not what we want for quantum mechanics.  Nonetheless, 
it is closely related to the quantum phase space. 

The Heisenberg subalgebra generated by $\{X_i, P_i, I\}$ is an invariant 
one. Note that by taking out the central charge generator $I$ one does 
not even have a subalgebra.  We start with the familiar coherent state 
representation 
\bea
e^{i\theta}\left|p^i,x^i\rra = U(p^i,x^i,\theta) \left|0\rra
\eea
where
\bea
U(p^i,x^i,\theta) \equiv 
e^{i\frac{x_i p^i}{2}} 
e^{i\theta\hat{I}} 
e^{-ix^i\hat{P}_i} 
e^{ip^i\hat{X}_i} 
= e^{i(p^i\hat{X}_i- x^i\hat{P}_i +\theta\hat{I})} \;,
\eea
and $\left|0\rra \equiv \left|0,0\rra$ is a fiducial normalized vector, 
$\hat{X}_i$ and $\hat{P}_i$ are representations of the generators $X_i$ 
and $P_i$ as Hermitian operators on the Hilbert space spanned by all of the
six parameter set of vectors $\left|p^i,x^i\rra$, and $\hat{I}$ is the identity 
operator representing the central generator $I$. Here, $(p^i,x^i,\theta)$ 
corresponds to a generic element of the (Heisenberg-Weyl) subgroup as  
\bea
{W}(p^i,x^i,\theta) = \exp i(p^iX_i- x^iP_i +\theta I)
\eea
with
\bea\label{ww}
{W}(p'^i, x'^i, \theta')  {W}(p^i, x^i, \theta)
= {W}\!\!\left(p'^i+p^i, x'^i+x^i, \theta'+\theta-\frac{x'_i p^i - p'_i x^i}{2} \right) \;.
\eea
where $x'_i p^i - p'_i x^i $  is the classical mechanical symplectic form \cite{K,D} . 
This is an infinite-dimensional unitary representation \cite{K,D}. This 
Hilbert space, or rather its projective counterpart, is the phase space for
the quantum mechanics.  The projective Hilbert space is, in fact, an 
infinite-dimensional symplectic manifold. Note that $p^i$ and $x^i$, as 
labels of the coherent states, correspond to expectation values, but not 
eigenvalues of the $\hat{P}_i$ and $\hat{X}_i$ observables. The coherent 
states give an overcomplete basis, with overlap given by
\bea\label{ol}
\lla p'^i,x'^i\right. \left|p^i,x^i\rra
&& \!\!\!\!= 
\exp\!\! \left[i\frac{x'_i p^i - p'_i x^i}{2} \right]
\exp\!\! \left[-\frac{(x'^i-x^i)(x'_i-x_i)+(p'^i-p^i)(p'_i-p_i)}{4}\right]
\nonumber \\[5pt]
&&\hspace{.5in}\xrightarrow{\quad p' \rightarrow p,  \;\; x' \rightarrow x \quad} 1 \;.
\eea
We also have
\bea
\lla p'^i,x'^i\right| \hat{X}_i \left|p^i,x^i\rra
&=& \frac{(x'_i+x_i)-i(p'_i-p_i)}{2}
\lla p'^i,x'^i\right. \left|p^i,x^i\rra \;,
\nonumber \\
\lla p'^i,x'^i\right| \hat{P}_i \left|p^i,x^i\rra
&=&  \frac{(p'_i+p_i)+i(x'_i-x_i)}{2}
\lla p'^i,x'^i\right. \left|p^i,x^i\rra  \;,
\label{xpo}
\eea
which are important results for our analysis below.

The above coset space is modeled on the Heisenberg-Weyl subgroup.  Explicitly, 
\bea
 \left(\begin{array}{cccc}
1 & 0 & 0 & p^i \\
0 & 1 & 0 & x^i \\
-\frac{1}{2}x_i  & \frac{1}{2}p_i  & 1 & \theta\\
0 & 0 & 0 & 1
\end{array}\right) \left(\begin{array}{cccc}
R^i_j & 0 & 0 & 0\\
0 & R^i_j & 0 & 0 \\
0 & 0  & 1 & 0\\
0 & 0 & 0 & 1
\end{array}\right)
=\left(\begin{array}{cccc}
R^i_j & 0 & 0 & p^i \\
0 & R^i_j & 0 &  x^i \\
-\frac{1}{2}x_i  R^i_j & \frac{1}{2}p_i R^i_j   & 1 & \theta\\
0 & 0 & 0 & 1
\end{array}\right) \;.
\eea
For fixed $(p^i,x^i,\theta)$, the above gives a generic element of the coset 
with the $R^i_j$ taken as elements of the $SO(3)$ subgroup. The fiducial vector 
$\left|0,0\rra$ corresponds to $(0,0,0,1)^t$ which is taken by any such coset 
onto $(p^i,x^i,\theta,1)^t$ corresponding to $e^{i\theta}\left|p^i,x^i\rra$. 
This illustrates explicitly the $U(p^i,x^i,\theta)$ action of the Heisenberg-Weyl
subgroup, and in fact also its extension to the full group on the Hilbert space,
as depicted on the coset space. Each transformation of the unitary 
representation sends a coherent state to another coherent state and hence 
its action can be depicted in the coset space with elements of the latter 
mapped to the coherent states. 

As inspired by the Galilean/Newtonian case, we can take a different coset space 
representation by factoring out an $ISO(3)$ subgroup generated by the $X_i$ 
and $J_{ij}$. The infinitesimal, or algebra, representation is then given as
\bea \label{qsc}
\left(\begin{array}{c}
dx^i  \\  d\theta \\    0
\end{array}\right) =
\left(\begin{array}{ccc}
 \omega^i_j &   0  &  \bar{x}^i\\
\bar{p}_j  & 0   &   \bar{\theta} \\   
 {0}   & {0} &  0 
\end{array}\right)
\left(\begin{array}{c}
x^j \\ \theta \\    1
\end{array}\right)
=\left(\begin{array}{c}
\omega^i_j  x^j + \bar{x}^i\\
  \bar{p}_j  x^j  +\bar{\theta} \\   0
\end{array}\right) \;.
\eea
The $(x^i, \theta)$ space is the quantum counterpart for the coset space 
that describes Newtonian (configuration) space. We can also construct a 
unitary representation whose relation to the coset is the same as the 
above for the phase space.  The $P_i$ and $I$ generators give group 
elements matching to the points in the coset space, and also generate 
an invariant subalgebra, which is however, trivial. This can also be seen 
in the corresponding group structure, i.e. by defining
\bea
W'(x^i,\theta)=\exp i(-x^iP_i+\theta I)\;,
\eea
we obtain
\bea 
W'(x'^i,\theta')W'(x^i,\theta)=W'(x'^i+x^i,\theta'+\theta)\;. 
\eea
We have a picture here very similar to the coherent state representation 
above with basis vectors labeled by the coset coordinates $x^i$ such that
\bea
e^{i\theta}\left|x^i\rra = U'(x^i,\theta) \left|0\rra
\eea
where
\bea
U'(x^i,\theta) \equiv e^{i\theta\hat{I}}
e^{-ix^i\hat{P}_i}  \;,
\eea
$\left|0\rra$ is the fiducial normalized vector, and $\hat{P}_i$ and $\hat{I}$ 
Hermitian operators on the Hilbert space spanned by all of the three parameter 
set of $\left|x^i\rra$ vectors. Much the same as before, $\hat{P}_i$ generates 
translations in $x^i$, while $\hat{I}$ is the identity operator effectively 
generating only a phase rotation of a vector on the Hilbert space spanned by all 
$\left|x^i\rra$. Following the coset action, we can see again the action of the 
unitary representation for the full group of $H_{\!\ssc R}(3)$. In particular, we 
see that 
\bea
e^{ip^i \hat{X}_i} e^{i\theta}\left|x^i\rra
=e^{i(p^i x_i+ \theta)} \left|x^i\rra \;, 
\eea
thus illustrating that the vectors $\left|x^i\rra$ are really the usual position 
eigenstates. The unitary representation constructed here from the coset space 
describing the quantum analog of the free particle configuration space, or 
physical space, is the configuration analog along the lines of the phase space
construction. It is however equivalent to that of the latter as a Hilbert space.

\section{Newtonian Limit from a Symmetry Contraction}
A naive way of interpreting the coset representations given above as quantum
analogs of the classical (configuration) space and phase space is suggested 
by simply replacing the generator $I$ by zero and dropping the variable
$\theta$ from consideration. A symmetry contraction, however, gives
a solid mathematical way to formulate the classical theory as an 
approximation to the quantum theory.
Consider the contraction \cite{c} of the above Lie algebra, given by the 
$k \to \infty$ limit under the rescaled generators $X_i^c=\frac{1}{k} X_i$ 
and $P_i^c=\frac{1}{k} P_i$. The $J$-$P^c$ and $J$-$X^c$ commutators are 
the same as those of $J$-$P$ and $J$-$X$; however, we have
\[
[X_i^c,P_j^c]=\frac{i}{k^2}\delta_{ij} I \quad \rightarrow \quad 0 \;,
\]
giving the commuting classical position and momentum. The contracted 
Lie algebra gives, with the $H$ generator included,  the Galilean relativity 
symmetry with a trivial central extension, in which $I$ is decoupled.
The symmetry contraction applied to the above representations also gives 
exactly the classical phase space, as well as Newtonian space, as we will see. 

The algebra element should first be written in terms of the rescaled 
generators as  $-i\left(\frac{1}{2}\omega^{ij}J_{ij}+\bar{p}^i_cX_i^c
-\bar{x}^i_cP_i^c+\bar{\theta}I\right)$. 
It is important to note that the parameters $\bar{p}^i_c =k\bar{p}^i$ 
and $\bar{x}^i_c =k\bar{x}^i$ are to be taken as finite even in the 
$k \to \infty$ limit. They are then parameters of the contracted algebra. 
The coset space of $(p^i,x^i,\theta)$  should be described in terms of 
$(p^i_c,x^i_c,\theta)$ with the representation rewritten as
\bea \label{psc}
\left(\begin{array}{c}
dp^i_c  \\ dx^i_c \\ d\theta \\    0
\end{array}\right) =
\left(\begin{array}{cccc}
 \omega^i_j &  0& 0  &  \bar{p}^i_c \\
0  & \omega^i_j &  0 &  \bar{x}^i_c  \\
-\frac{1}{2k^2}\bar{x}_{cj}   & \frac{1}{2k^2}\bar{p}_{cj}   &  0 & \bar{\theta} \\
 {0}   & {0} &  0 & 0
\end{array}\right)
\left(\begin{array}{c}
p^j_c  \\ x^j_c   \\ \theta \\    1
\end{array}\right)
=\left(\begin{array}{c}
\omega^i_j  p^j_c  + \bar{p}^i_c \\
\omega^i_j  x^j_c  + \bar{x}^i_c  \\ 
 \frac{1}{2k^2}(  \bar{p}_{cj}    x^j_c-\bar{x}_{cj} p^j_c  )   + \bar{\theta} \\   0
\end{array}\right) \;.
\eea
This gives only $d\theta=\bar{\theta}$ in the limit; hence, $\theta$ 
becomes an absolute parameter not affected by the transformations, 
except its own translation generated by $I$. Note that $dp^i_c$ and  $dx^i_c$ 
are also $\bar{\theta}$-independent.  This reflects exactly what we mean when 
saying that $I$ decouples. The $\bar{\theta}$ parameter has nothing to do 
with anything else any more. It may as well simply be dropped from consideration.  
The  $(p^i_c, x^i_c)$ space is exactly the classical phase space. We have a 
parallel result for the other coset; explicitly,
\bea \label{sc}
\left(\begin{array}{c}
dx^i_c  \\  d\theta \\    0
\end{array}\right) =
\left(\begin{array}{ccc}
 \omega^i_j &   0  &  \bar{x}^i_c \\
\frac{1}{k^2}\bar{p}_{cj}  & 0   &   \bar{\theta} \\
 {0}   & {0} &  0 
\end{array}\right)
\left(\begin{array}{c}
x^j_c  \\ \theta \\    1
\end{array}\right)
=\left(\begin{array}{c}
\omega^i_j  x^j_c + \bar{x}^i_c\\
  \frac{1}{k^2}\bar{p}_{cj}  x^j_c   +\bar{\theta} \\   0
\end{array}\right) \;.
\eea
giving only $dx^i_c=\omega^i_j  x^j_c + \bar{x}^i_c$ and $d\theta=\bar{\theta}$.

We can also apply the symmetry contraction to the unitary representations 
given on the above Hilbert space(s).  We first look at the latter as a representation of
the algebra of observables, based on $\hat{X}_i^c$ and $\hat{P}_i^c$ (and 
$\hat{I}$) at finite $k$. The set of $\left|p^i,x^i\rra$ states should be
re-labeled as $\left|\tilde{p}_i^c,\tilde{x}_i^c\rra$, with the $\tilde{p}_i^c$
and $\tilde{x}_i^c$ characterizing the expectation values of $\hat{X}_i^c$ 
and $\hat{P}_i^c$. Note that $\tilde{p}_i^c$ and $\tilde{x}_i^c$ do not directly 
correspond to the $p_c^i$ and $x_c^i$ above. From Eqs. (\ref{ol}) and (\ref{xpo}) then,  
we have
\bea
\lla \tilde{p}'^c_{i},\tilde{x}'^c_{i} \right| \hat{X}_i^c \left|\tilde{p}_i^c,\tilde{x}_i^c\rra
&=& \frac{(\tilde{x}'^c_{i}+\tilde{x}^c_{i})-i(\tilde{p}'^c_{i}-\tilde{p}^c_{i})}{2}
\lla \tilde{p}'^c_{i},\tilde{x}'^c_{i} \right.  \left|\tilde{p}_i^c,\tilde{x}_i^c\rra \;,
\nonumber \\
\lla \tilde{p}'^c_{i},\tilde{x}'^c_{i}\right| \hat{P}_i^c \left|\tilde{p}_i^c,\tilde{x}_i^c\rra
&=& \frac{(\tilde{p}'^c_{i}+\tilde{p}^c_{i})+i(\tilde{x}'^c_{i}-\tilde{x}^c_{i})}{2}
\lla \tilde{p}'^c_{i},\tilde{x}'^c_{i} \right.  \left|\tilde{p}_i^c,\tilde{x}_i^c\rra \;,
\eea
where the state overlap has the second, real and negative exponential 
factor, written in terms of $\tilde{p}'^c_{i}$, $\tilde{x}'^c_{i}$, $\tilde{p}_i^c$
and $\tilde{x}_i^c$, proportional to $k^2$. This therefore gives a vanishing result in the 
contraction limit, so long as the coherent states are not the same. Thus, 
we can see that $\hat{P}^c_i$ and $\hat{X}^c_i$ are diagonal 
on $\left|\tilde{p}_i^c,\tilde{x}_i^c\rra$ with $\tilde{p}_i^c$ and 
$\tilde{x}_i^c$ as eigenvalues.
The Hilbert space, as a representation for the Heisenberg-Weyl symmetry
and that of the algebra of observables described as functions (or polynomials) 
of $\hat{P}^c_i$ and $\hat{X}^c_i$, is therefore reducible. It reduces to a 
direct sum of one-dimensional representations of the ray spaces of each 
$\left|\tilde{p}_i^c,\tilde{x}_i^c\rra$. That is to say, the only admissible states 
are the exact coherent states, and not any linear combinations. These are really
the classical states, though we are not used to describing classical mechanics 
in the Hilbert space language. Actually, this kind of description has been 
available for a long time \cite{cHsp}. The latter may be particularly useful in
establishing the more involved dynamical picture of what we discussed here. 
Note that $\hat{H}^c$ and $\hat{J}_{ij}^c$ as classical observables would also 
be diagonal on $\left|\tilde{p}_i^c,\tilde{x}_i^c\rra$. However, in getting the 
contracted symmetry algebra, the generators $J_{ij}$ (and $H$) are not to be 
rescaled by $k$, and maintain all their nonzero commutators. The 
transformations they generate still take one state to another, as they should 
in the classical picture. As they always take a coherent state to a coherent 
state anyway, they do not support linear combinations either. The set of
coherent states essentially gives just the classical coset/phase space. Readers 
may find interest in an explicit expression of the generator for dynamical/time 
evolution on the Hilbert space of $\left|\tilde{p}_i^c,\tilde{x}_i^c\rra$ 
states in terms of the classical Hamiltonian \cite{cHsp}.  

The story for the contraction of the Hilbert space as the quantum configuration space is 
somewhat less obvious. The basis vectors are eigenvectors of transformations
in the $W'$ group. But we know that this space serves as a representation for 
the Heisenberg algebra, and hence the algebra of observables, on which the 
action of the $\hat{P}_i$ operators for the momentum observables have
eigenstates being linear combinations of the basis $\left| x^i \rra$ states.
It is exactly for such considerations of momentum-dependent observables that 
one needs to go beyond the coset to the full Hilbert space. At the contraction 
limit, however, the Heisenberg algebra is trivialized. $\hat{P}_i^c$ commuting 
with $\hat{X}_i^c$ means that they have to share the same eigenvectors 
$\left|\tilde{x}_i^c\rra$, now labeled by the $\hat{X}_i^c$ eigenvalues.
Again the Hilbert space as a representation for the algebra of observables 
reduces and only the latter vectors are relevant, not linear combinations 
or even the phases. The result is the Newtonian three-dimensional space.
 
Readers should have realized that our rescaling parameter $k$ for 
the implementation of the symmetry contraction corresponds to 
$\frac{1}{\sqrt{\hbar}}$, so that the contraction is really the $\hbar \to 0$ 
limit. The latter of course corresponds to taking the classical approximation. 
In fact, the quantum symmetry algebra is quite commonly written with an 
$\hbar$ within each commutator. Our version first refers to the natural 
quantum units of $\hbar=1$, in which the $J_{ij}$ are dimensionless.  The 
contraction limit is obtained as described, which is the same as taking 
{\em only} the $\hbar$ in the $\hat{X}^c$-$\hat{P}^c$ commutator to zero. 
Otherwise, all commutators would be killed. If one is taking the algebra as 
describing relations among the classical observables, this is great; however,
considering it as the relativity symmetry algebra, this is a disaster. The 
algebra of observables is really not the one for the relativity symmetry, but 
rather the algebra of functions of $\hat{X}^c$ and $\hat{P}^c$, and as such
a specific representation of the relativity algebra. Nevertheless, we still need 
to re-introduce the nonzero $\hbar$ in the the rest of the relativity algebra 
to have $J_{ij}$ (and $H$) being described in the classical units, if we want 
to match them to the observables $\hat{J}_{ij}^c$ and $\hat{H}^c$. Generators 
of the relativity symmetry are {\em not} to be identified with the operators 
representing the observables in the Hilbert space picture of classical 
mechanics \cite{cHsp,L}. The contraction is not concerned with the units. 
In the classical picture after the contraction, it is no longer un-natural to 
have units for position and momentum chosen as independent,
hence their product having a nontrivial unit. That unit would have 
fundamental significance in telling when the classical theory is a good 
approximation to the better quantum theory.  The contraction is a 
mathematical procedure for getting the approximate theory characterized 
by a small scale \cite{c}. The classical scale is the one which is small
compared to the contraction parameter $k$, hence with smallness 
described by the $\hbar$ value. $\hbar$ serves as the fundamental 
unit with which we re-express physical quantities.

The coset space pictures at least illustrate well that quantum space 
is different from Newtonian space in much the same way as the 
quantum phase space is different from the classical one. The analysis 
of the (equivalent) infinite-dimensional unitary representations and 
their reductions upon the symmetry contraction gives the full, 
solid results. 

\section{Concluding Remarks}
We physicists should not endow a vague common sense concept like 
physical space with any particular mathematical model as a given. We are 
supposed to learn from experiments what constitutes a good/correct
theoretical/mathematical model of any physical concept, and physical space 
should not be an exception. We have by now roughly a century of experimental 
results saying that the classical/Newtonian model of physical space does 
not serve this purpose so well, especially not as the configuration space of  
quantum particle motion. We should not be reluctant to modify it.  What 
could the notion of (classical/Newtonian) space, described in any inertial 
frame, be other than the configuration space of (free) particle motion under 
arbitrary initial conditions? What kind of coordinates would be more natural 
for space besides the $q^i$ variables acting as the angle coordinates, with 
$p^i$ as action coordinates, for free particle motion as described on the 
phase space? Looking at physical space as it can possibly be understood 
from practical physics, the space of  all $q^i$ values as the configuration 
variables is essentially the only picture we should have, so long as 
nonrelativistic `particle' mechanics, classical or quantum, are concerned. 

It is known that the projective Hilbert space, as the true quantum phase 
space, is an infinite-dimensional symplectic manifold.  An expansion of 
a state in terms of an orthonormal basis in the form
$\left| \phi \rra =\sum (q_n +ip_n)\left| n \rra$  gives $q_n$ and $p_n$ 
as a set of real homogeneous coordinates  of the projective space on 
which the Schr\"odinger equation is equivalent to the set of Hamilton 
equations of motion for $q_n$ and $p_n$ as pairs of configuration and 
momentum variables with Hamiltonian function 
$H(p_n,q_n)=\frac{2}{\hbar} \lla \phi \big|\hat{H}\big| \phi \rra$. It suggests 
thinking about a Lagrangian submanifold, like the space of the $q_n$, as 
the quantum configuration space. One can also take the real and imaginary
part of the values of a wavefunction at the various points (of the 
classical space model) as a similar set of symplectic coordinates. 
However, our perspective of the quantum relativity symmetry has a 
complex phase rotation of the state generated by the X-P commutator 
which mixes the configuration and momentum coordinates. Hence, unlike
the classical case, the position/configuration space and the momentum
space are no longer irreducible components of the relativity symmetry.
The quantum phase space is an irreducible representation, though the
classical one is reducible. We get a quantum (position) space model
that is equivalent to the phase space model. The projective Hilbert 
space is also to be a  Kahl\"er manifold \cite{B,km}, and hence 
has a natural metric, though the latter notion may not be 
feasible on a generic symplectic manifold.

The analysis in this article is simple and straightforward, with results hardly 
totally new or unexpected for the phase space picture. What is new and 
important is the way they are pieced together consistently to illustrate the 
basic perspective; and that the application of the latter suggests looking at 
familiar notions in quantum physics in a very different way. In particular,
it gives a picture of the not quite discussed notion of the configuration 
space in quantum mechanics as a model of physical space beyond the usual 
one, which is nothing but the Newtonian model. This is the first step in 
justifying a new perspective regarding (quantum) physical space, the adoption 
of which may also help clarify some issues in quantum physics and beyond.

Symmetry is the single most important organizing principle in the theory 
of modern physics. What we performed in the above analysis is an attempt 
to see how the fundamental symmetry of something like free particle motion 
informs us about the nature of the phase space, configuration space, and 
hence our physical space. These types of symmetries are relativity symmetries. 
Different fundamental theories have different relativity symmetries, which 
correspond to different pictures of physical space and time, just like 
Einsteinian (special) relativity gives a Minkowski spacetime. In fact, the 
mathematical relation of the latter to the Newtonian one can be described 
exactly using the corresponding coset space picture as representations of the 
relativity symmetries through the symmetry contraction with $c$ as the 
parameter \cite{EtN,060}. The above symmetry contraction is really the 
necessary, proper, and quite subtle, mathematical way to describe the 
Newtonian limit as an approximation to the better Einsteinian or quantum 
theory. We give the analogous mathematical description of the quantum to
classical case here and use it to illustrate a picture of quantum space. In 
this case $\hbar$, or rather $\frac{1}{\sqrt\hbar}$, takes the place of $c$.
Neither $\hbar$ nor $\frac{1}{c}$ is really zero: nonzero values of both
are key fundamental constants.  The symmetry contraction limit provides
the necessary subtle approach to successfully describe the Newtonian 
approximation.

Given the basic perspective of looking for a picture of quantum space
as described by the symmetry structure of the theory instead of the
corresponding classical notion, the considerations and analysis presented 
here is necessarily simple and somewhat naive. As such, it is definitely not 
the 'final' answer in the general setting of quantum physics. Invariance under
Einsteinian special relativity, for example, has not been incorporated. Our
key point of interest here is exactly in showing how this basic perspective 
provides us with a notion of quantum space(time) beyond classical space(time), 
yet giving rise to the latter when the proper limit is taken, even for the 
simplest, ordinary and conventional theory of quantum mechanics 
without any extra assumptions. Hence, we are not interested here in 
putting in extra notions beyond the bare minimum, no matter how natural
one may argue for them to have a part in quantum physics. In fact, the 
basic perspective, we believe, can take us much beyond the simple 
results in this article.  Our study on quantum spacetime, given by the work
presented here, is therefore necessarily incomplete. Moreover, our 
discussion has been entirely restricted to kinematics - analysis of the 
full dynamical picture will be given in a separate publication \cite{070}.
There are two main reasons for separation the two. Conceptually, as
seen in the Newtonian example, the constructions of the notion of
particle configuration space and phase space, as well as that of physical 
space, require only kinematical considerations. Besides this, as to be 
reported in \cite{070}, the dynamical picture should firstly be considered 
as one on the algebra of observables rather than the configuration space 
or phase space. Otherwise, the Schr\"odinger equation applied to the 
set of coherent states is known to be equivalent to the classical
dynamics on the states taken as classical ones. That is all that is relevant 
so long as the dynamics of the pure states of the quantum Hilbert 
space is concerned. A further source of incompleteness lies in the 
fact that field theory issues are not discussed here either. Note that 
practical field theories are either quantum or at least (Einstein) 
relativistic. It goes without saying that we have the big task at hand of 
extending this framework to the fully deformed/stabilized fundamental
quantum relativity. We hope that the simple analysis here can 
help make our basic perspective more accessible to general readers, 
beyond those who have more experience with spacetime physics 
and the foundations of quantum mechanics, as well as new developments 
in these areas.

Our group has worked on a notion of a quantum relativity for deep 
microscopic quantum spacetime \cite{030}, from much the same 
theoretical perspective as that which lies behind the current analysis. The 
basic starting point there is the old idea of relativity deformations \cite{rd} 
to which contraction of the relativity symmetries is the reverse process,
so long as one stays within the Lie group/algebra framework.
While we have presented some picture of the physics from a sequence of 
contractions \cite{060}, we are currently working on the details of the
descriptions of an alternative contraction scheme, via an approach that 
naturally incorporates symmetries like $H_{\!\ssc R}(3)$ and 
$\tilde{G}(3)$. The results here are really part of that work. The `final' 
symmetry is considered to have non-commuting $X_i$ and $P_i$
 \cite{030}, to which no real number picture of spacetime is expected 
to work. Within the domain of simple quantum mechanics investigated 
here, the physical space picture still looks like a real manifold, albeit of 
infinite dimension. The results here may also serve as the crucial first 
link from the bottom-up to any theoretical model of spacetime beyond 
the level of simple quantum mechanics.

A fair question is if it is too conservative to stay within the Lie 
group/algebra framework. While we sure encourage other alternative
bold approaches within the deformed relativity picture, what we want
to emphasize is that our chosen framework is a very powerful one.  
The $H_{\!\ssc R}(3)$, or $\tilde{G}(3)$, group obviously corresponds to an
observable algebra which is quantum/noncommutative. In fact, the 
latter is more or less just the group $C^*$ algebra \cite{070}, which is 
a completion of the group algebra \cite{gca}. The quantum Hilbert space
is naturally a cyclic irreducible representation of the algebra 
corresponding to its space of pure states  \cite{070}. The theory of
noncommutative geometry \cite{ncg} says any (noncommutative)
algebra has a matching topological/geometric space which we see 
as essentially the projective Hilbert space in our case. It is then
indeed quite plausible that the picture of relativity symmetries as
Lie groups is a good enough starting point to formulate the
noncommutative geometries of quantum spacetime. Again, the 
representation contraction picture gives the setting to build 
kinematic and dynamic models which can be systematically traced  
back to those of well-known physics.

To look at the dynamical picture at the quantum level under a formulation 
completely in line with our approach here is mathematically involved. The
Weyl-Wigner-Groenewold-Moyal formalism has to first be rewritten
with the coherent state basis or wavefunctions $\lla p^i,x^i | \phi \rra$
as the starting point and fully matches to a representation picture of
the group $C^*$ algebra, though restriction of the latter to that of the 
Heisenberg-Weyl subgroup is good enough. Thanks to the semidirect 
product structure, a  representation of the subgroup  and its $C^*$ 
algebra serves as a representation the full group ($C^*$ algebra) in
which  elements beyond the subgroup  act as inner automorhisms 
\cite{070}. The observable algebra is the representation of the group
$C^*$ algebra. Naively summarized, so long as the contraction to the
classical limit is concerned, it is just the reverse of the standard 
deformation quantization in the $\hbar \to 0$ limit.  A generator of
the full relativity symmetry group $G_{\!s}$ is represented by a function
 $G_{\!s}(\hat{P}_i,\hat{X}_i)$ with 
$G_{\!s}(\hat{P}_i,\hat{X}_i)\star= G_{\!s}(p_i \star, x_i \star)$, as
an operator acting on the Hilbert space of wavefunctions and
the observable algebra itself, in which $\star$ is the standard Moyal
star product. The latter action is the left regular representation of the
algebra on itself, and there is a corresponding right action. However, the 
corresponding automorphisms of the observable algebra which match with
the unitary transformations on the Hilbert space are really generated by
the difference of the left and the right action. This can be written 
as $\{G_{\!s}(p_i, x_i), \cdot\}_\star$, i.e. in terms of the Moyal bracket. 
In the $\hbar \to 0$ limit, formulated here as the $k\to \infty$ limit
as described above, the $G_{\!s}(p_i \star, x_i \star)$ action reduces
to the classical multiplicative action of $G_{\!s}(p_i, x_i)$, as all classical 
observables commutes. The generators for the automorphisms as
symmetry transformations in the Heisenberg picture, however, 
reduce to the classical Liouville operator; hence giving the Poisson
algebra structure. Time evolution is just the symmetry transformation
generated by the Hamiltonian operator/function. Hence, one retrieves
classical dynamics. The separate notion of a function as a multiplicative
operator and its corresponding Liouville operator have been studied 
in the Koopman-von Neumann formalism \cite{L}, which is really a
Hilbert space picture for the mixed states. All of this can be retrieved 
as the contraction limit \cite{070}, except the naive Schr\"odinger
picture of dynamics. We have seen above that the quantum Hilbert
space of pure states reduces to essentially that of the classical phase
space. In the Hilbert space picture, the classical pure states are 
essentially disconnected vectors/rays. It is then no surprise at all
that one does not have a Schr\"odinger dynamics for the classical
pure states as the contraction limit. Classical dynamics is really one
of the Heisenberg picture. For details, readers are referred to the
companion paper \cite{070}.

Somewhat after the posting of the first version of this paper, another study 
of the notion of  model for the physical space behind quantum mechanics
 \cite{FLM} came up. The approach there has nothing to do with the theme
of relativity symmetry contraction/deformation here. Nevertheless, it
may be in the interest of the readers for us to give a comparison between
their approach and ours in this paper and beyond. As stated with emphasis in 
their introductory section, Ref.\cite{FLM} is focused on {\em``quantum systems 
with a built-in length scale."} We sure share the idea that some fundamental 
scale(s) being built into the basic formulation would indeed be an important 
part of any theory of deep microscopic quantum spacetime. We have 
the relativity symmetries for simple quantum mechanics and classical
Newtonian mechanics as retrieved from the proper (contraction) limits 
of the such a quantum relativity symmetry  \cite{030}. The limits provide 
the setting within which the fundamental scales can be neglected.
No matter how natural the idea of having fundamental quantum scales 
may sound to many of us, saying that it is a part of the ordinary 
(formulation of) quantum mechanics may really be pushing it too much. 
Our analysis here is particularly interested in developing a notion of quantum 
space without putting such kinds of extra theoretical structure into ordinary 
quantum mechanics. Ref.\cite{FLM} illustrates how their notion of 
modular space-time is arguably a natural part of quantum mechanics 
with a fundamental (length) scale, which is certainly of great interest. 
It is, however, beyond the setting of ordinary quantum mechanics. 
There is however an important difference between our perspectives 
on quantum spacetime in general.  The ``{\em point of view that any choice 
of a maximally commutative $*$-subalgebra of the Heisenberg algebra can be 
thought of as defining our concept of quantum Euclidean space}" \cite{FLM} 
is to be contrasted against our point of view that the full quantum 
noncommutative algebra of observables can be thought of as defining 
a concept of quantum space(time), which is generally noncommutative
\cite{ncg}. As discussed in \cite{030}, fundamental scales are supposed to 
characterize noncommutativity of the classical notion of spacetime 
coordinates as well as and momentum coordinates. This perspective is the 
key that gives - even in the current (limited) setting without fundamental 
scales - a notion of quantum space beyond the classical. The notion of 
``quantum Euclidean space" in Ref.\cite{FLM} will likely be retrievable 
from proper limits of our idea of noncommutative quantum spacetime
from the full relativity symmetry with fundamental scales incorporated
\footnote{It is interesting to note the following: the fundamental quantum
relativity symmetry of \cite{030} can be written as
\bea
&& 
[X_\mu, X_\nu] =i M_{\mu\nu}  \;, \qquad
[P_\mu, P_\nu] =-i M_{\mu\nu} \;, 
\nonumber \\ &&
[X_\mu, P_\nu]=i\eta_{\mu\nu} F \;, \qquad
[X_\mu, F]=-iP_\mu \;, \qquad
[P_\mu, F]=-iX_\mu \;,
\nonumber 
\eea
$\eta_{\mu\nu}=\mbox{diag}\{-1,1,1,1\}$,
with all fundamental scales taken as unity.
On an eigenspace of $M_{\mu\nu} $ and $F$ of integral eigenvalues,
as a representation space, the set of $e^{2\pi i X_\mu}$
and $e^{i P_\mu}$ behaves like the commuting set of $U$ and $V$
of the ``Heisenberg group" discussed in the modular picture of \cite{FLM}.
},
which is still to be constructed.

We have not touched on the measurement problem so far. A couple of 
comments on this issue are in order. To the extent that we do not have any 
dynamical theory to describe a measurement process \cite{deco}, our leaving 
such issues on the sideline is justified. We sure do not see the quantum 
space picture here as, in any sense, `final', and we do not aim at describing 
measurements. We want to note, however, that most if not all, discussions 
about measurements are really about classical measurements, as Bohr did a 
good job in elaborating. They are about extracting pieces of classical information,
as represented by numbers, from a quantum system. It is not surprising that 
the nature of the information/physical attributes of the system being quantum 
does not fit in well with such measurements. If the quantum position is to 
be described by infinitely many real numbers, our decision to `get' 
one or three real numbers reading to the so-called probabilistic 
results.  Only statistics from many such measurements can give a better
approximation of those infinite coordinate values. Actually, we essentially 
only obtain values of any measurements by comparison. For example, 
position or distance between two positions is measured by comparing
it to a length standard, admitting some uncertainty. The nature of that
`ratio' being a piece of classical information, a real number, is never more 
than a mathematical model or an assumption. With development of quantum 
information theory, physicists in the future may be proficient in handling 
quantum information and true quantum measurements may then be the rule, 
rather than the exception. We would like to advance the notion of 
measurements as possibly extracting quantum, non-real-number, information 
from a system which describes some of its properties. Even the idea of a  
`definite' position in physical quantum space may plausibly be useful for 
that kind of position information. However, we are certainly not defending the 
classical notion of being able to extract full information about a dynamical 
state without disturbing it at all. It is not our intent either to take a 
stand in that kind of philosophical debate about realism here, which we see 
as beyond, and not at all necessary to, the study of physics.

\acknowledgements
The authors are partially supported by research grants 
NSC 102-2112-M-008-007-MY3 and MOST 105-2112-M-008-017-
from the MOST of Taiwan.
We thank N. Gresnigt, P.-M. Ho,  and H.S. Yang
for helpful comments on the manuscript. \\

\end{document}